# Dynamic transition from Mott-like to metal-like state of the vortex lattice in a superconducting film with a periodic array of holes

*Indranil Roy$^a$, Prashant Chauhan$^b$, Harkirat Singh$^a$, Sanjeev Kumar$^b$, John Jesudasan$^a$, Pradnya Parab$^b$, Rajdeep Sensarma$^a$, Sangita Bose$^{b*}$ and Pratap Raychaudhuri$^{a†}$*

$^a$ Tata Institute of Fundamental Research, Homi Bhabha Road, Colaba, Mumbai 400005, India.
$^b$ UM-DAE Center for Excellence in Basic Sciences, University of Mumbai, Vidhyanagari Campus, Mumbai-400098, India.

We show that under an a.c. magnetic field excitation the vortex lattice in a superconductor with periodic array of holes can undergo a transition from a Mott-like state where each vortex is localized in a hole, to a metal-like state where the vortices get delocalized. The vortex dynamics is studied through the magnetic shielding response which is measured using a low frequency two-coil mutual inductance technique on a disordered superconducting NbN film having periodic array of holes. We observe that the shielding response of the vortex state is strongly dependent on the amplitude of the a.c. magnetic excitation. At low amplitude the shielding response varies smoothly with excitation amplitude, corresponding to elastic deformation of the vortex lattice. However, above a threshold value of excitation the response shows a series of sharp jumps, signaling the onset of the Mott to metal transition. Quantitative analysis reveals that this is a collective phenomenon which depends on the filling fraction of vortices in the antidot lattice.

$^†$e-mail: pratap@tifr.res.in
$^*$e-mail: sangita.bose@gmail.com

## I. Introduction

Understanding the vortex dynamics in strongly pinned Type II superconductors is of paramount importance both from fundamental and practical point of view. On the one hand the vortex lattice (VL) provides a versatile model system to understand the interplay of interaction and pinning in a periodic solid[1,2]. On the other hand, the collective dynamics of the VL determines the critical current, which dictates the limit of the applicability of the material in low loss devices. Of particular interest are superconductors with periodic array of holes, i.e. antidot arrays, which act as strong pinning centers for the flux lines[3,4,5,6,7,8]. Here, besides the overall confining potential created by the shielding supercurrent, each vortex experiences two kinds of forces: The nearly field independent confining potential of the antidot and the field dependent confining potential caused by the repulsive interaction of surrounding vortices. At matching fields, corresponding to integer number of vortices in each antidot, each vortex gets tightly confined in a cage formed by the neighboring vortices. This gives rise to pronounced oscillatory response of the superconductor with magnetic field, known as vortex matching effect (VME), which manifests as oscillations in magnetoresistance close to the superconducting transition temperature ($T_c$), with period corresponding to the first matching field ($H_m$). It has been suggested that when thermal fluctuations are not large enough to overcome the combined confinement caused by the antidot and the intervortex interactions, the vortex state close to matching field mimics that of a Mott insulator[9], where each vortex is localized around the antidot. Early evidences of the vortex Mott state in superconductors with periodic pinning[10,11] was further confirmed from detailed measurement of the compressibility of the vortex lattice[12]. Recently, transition of this vortex Mott insulator state to a metallic state, where vortices become delocalized, as a function of magnetic field or induced by an external current[13] has also been reported.

In this paper, we investigate the dynamic response of the VL in a disordered superconducting NbN film with antidot array, when subjected to an a.c. magnetic field with varying amplitude. We focus on the low temperature limit ($T \ll T_c$) where thermally excited vortex motion does not play a significant role. Usual transport measurements, i.e. resistivity and critical current, conventionally used to study VME cannot be used here, since they are restricted to temperatures[14] very close to $T_c$. To overcome this difficulty, we developed a "two-coil" mutual inductance technique[15] which can extend these measurements well below $T_c$. Here the superconducting film with antidot array is sandwiched between a quadrupolar primary coil and a dipolar pick-up coil. When a small a.c. current ($I_{ac}$) is passed through the primary, the flux tubes undergo oscillatory motion about their equilibrium position, thereby inducing a voltage in the secondary. The VME reflects in periodic modulation of the magnetic shielding response. This is measured from the real and imaginary part of the mutual inductance, $M'^{('')}$, which are related to the compressibility of the VL and the dissipation inside the superconductor respectively. Here, we study the shielding response as a function of $I_{ac}$, which is proportional to the a.c. magnetic field. The central result of this paper is that as $I_{ac}$ is gradually increased, the VL undergoes a transition from a Mott-like to a metal-like state, at a characteristic $I_{ac}$ value which is much smaller than the depinning threshold for an isolated vortex from the antidot, showing that this is a collective phenomenon dominated by vortex-vortex interactions. Furthermore, the $I_{ac}$ at which this transition happens is strongly dependent on the filling fraction of the vortex lattice and is lowest at $H_m$ when the VL is most rigid. Interestingly this is opposite to the behavior inferred from critical current measurements[5] close to $T_c$.

**II.    Sample details and measurement scheme**

*Sample.* The sample used in this study consists of a 3 mm diameter disordered NbN film with thickness, $t \sim 25$ nm, grown using reactive magnetron sputtering on nanoporous anodic alumina template (AAM). Details of synthesis and characterization have been reported earlier[16]. The condition for observing VME is that the inter-vortex repulsive force should be a non-negligible fraction of individual pinning force by the antidot. While close to $T_c$ the divergence of the Ginzburg-Landau coherence ($\xi_{GL}$) length ensures that this condition is easily satisfied[15], the observation of VME at $T \ll T_c$ requires careful tuning of material parameters. In NbN, the introduction of disorder (in the form Nb vacancies) decreases the strength of individual pinning through an increase[17] in $\xi_{GL}$ and a decrease in the superfluid density[18,19], making this condition achievable down to low temperatures. Consequently, we perform our measurements on a disordered NbN film with $T_c \sim 4.21$ K, for which vortex matching effects are observed down to the lowest temperature of our measurement (1.45 K). Scanning electron micrographs (SEM) of the film (Fig. 1(a)) obtained after deposition show pores with average diameter, $d = 46$ nm, nominally arranged in a hexagonal pattern with average lattice constant $a = 107$ nm. However, the hexagonal ordering is not perfect. 2-D Fourier transform of different areas[20] of the SEM image show six symmetric diffraction spots, but misoriented with respect to each other (right panels in Fig. 1(a)). This shows that, in addition to imperfection in the position of individual holes the nanopores also form domains, misoriented with respect to each other, similar to a polycrystal.

*Measurement.* The schematics of our two-coil mutual inductance setup is shown in Figure 1(b). The quadrupolar primary coil has 28 turns with the upper half wound in one direction and the lower half wound in the opposite direction, producing a peak field of 7 mOe/mA. The dipolar secondary coil consisting of 120 turns wound in 4 layers. To measure the mutual inductance a small sinusoidal excitation current ($I_{ac} \sim 1$-15 mA) with frequency $f = 31$ kHz is passed through the primary and the in phase ($V'$) and out of phase voltage ($V''$) induced in the secondary is measured using a lock-in amplifier. The real and imaginary part of mutual inductance are defined as, $M'^{('')}=V'^{('')}/(2\pi I_{ac}f)$, such that $M'$ and $M''$ correspond to the inductive and dissipative response of the sample. The two coil assembly is placed in a cryostat fitted with a superconducting solenoid capable of generating d.c. magnetic field ($H \perp$ film plane) up to 90 kOe.

To understand what the quantities $M'$ and $M''$ relate to, we note that the magnetic field produced by the primary coil will generate a *coarse grained averaged* time varying circulating supercurrent of the form, $\boldsymbol{J}_S^{ac}(\boldsymbol{r},t) = J_S^{ac}(r)\sin(\omega t)\,\widehat{\phi}$ in the superconducting film (Fig. 1(c)), which from symmetry considerations will be zero at the center of the sample and increase towards the periphery. (Here we ignore the pattern of supercurrent on a shorter length scale due to the presence of the antidots.) This will apply a radial oscillatory force on each vortex given by, $\boldsymbol{F}_{ac} = \boldsymbol{J}_S^{ac} \times \hat{z}\Phi_0 = J_S^{ac}(r)\sin\omega t\,\Phi_0\hat{r}$ causing each vortex to oscillate about its equilibrium position. Thus as a first approximation the effect of the a.c. excitation is to produce an oscillatory compaction and rarefaction of the VL with the vortex at the centre of the sample remaining in its equilibrium position (Fig. 1(d)). This is equivalent to a periodic change in the flux density which will produce an induced voltage in the secondary coil. Thus $M'$ is a measure of the compressibility of the VL. The amplitude of the imaginary part, $-M''$, on the other hand is related to the dissipation in the superconductor. At small a.c. excitations, when each vortex undergoes small oscillations about its equilibrium position, this dissipation arises from Bardeen-Stephen loss inside the superconductor surrounding the antidot. However, when the excitation is large enough to induce vortex hopping from one antidot to the next, the dissipation comes predominantly from the $2\pi$ phase slips in the intervening superconductor separating the two antidots.

## III. Results and discussion

Figure 2 shows the *M'* - *H* scans recorded at different temperatures between 3 K and 1.45 K with $I_{ac} \sim 5$ mA. *M'* shows dip at multiples of 2.37 kOe consistent with expected matching field, $H_M = \frac{2\Phi_0}{\sqrt{3}a^2}$, where $\Phi_0$ is the flux quantum. The dips in *M'* show that the VL become more rigid at the matching fields consistent with the description of a vortex Mott state[9,21]. Decreasing the temperature has two effects: First, the overall magnetic screening response increases and second, the minima in *M'* at matching fields become less pronounced. At a qualitative level, both these effects can be understood from the temperature variation of the coherence length, $\xi_{GL}$, and the magnetic penetration depth, $\lambda$, of the superconductor as outlined in ref. 15. The first effect can be understood from the fact that the pinning potential and therefore the individual pinning force, $F_p \sim \frac{\Phi_0^2}{4\pi\mu_o\Lambda\xi_{GL}}$, where $\Lambda (= 2\lambda^2/t)$ is the Pearl penetration depth $\mu_0$ is the vacuum permeability. Assuming the dirty Bardeen-Cooper-Schrieffer (BCS) relation, $\frac{\lambda^{-2}(T)}{\lambda^{-2}(0)} = \frac{\Delta(T)}{\Delta(0)} tanh\left(\frac{\Delta(T)}{2k_BT}\right)$ (where, $\Delta$ is the superconducting energy gap and $k_B$ is the Boltzman constant) $1/\lambda^2$ increases by 30% when we cool the sample from 3K to 1.45 K. Thus, at low temperatures, the vortex lattice becomes more strongly pinned thereby increasing the magnetic screening. On the other hand, the amplitude of oscillation in the screening response is governed by the inter-vortex repulsive force $F_I$, which renders the vortex lattice less compressible at matching fields when each vortex is surrounded by a vortex at each neighboring site. Since the repulsive force between two vortices in neighboring antidot is given by $F_I \sim \frac{\Phi_0^2}{4\pi\mu_o\Lambda a}$, $F_I/F_p \sim (\xi_{GL}/a)$ [15]. From the usual G-L relation $\xi_{GL} \propto \left(1-\frac{T}{T_c}\right)^{-1/2}$, we expect this ratio to decrease by 30% in the same temperature range, thereby making the oscillations less pronounced. We observe that at 1.45 K only the first matching field is clearly visible. Thus in the rest of the paper we concentrate primarily on the behavior around the first matching field.

The central result of this paper, namely, the screening response as a function of $I_{ac}$ (for constant $H$) is shown in Fig.3. Fig. 3(a) shows $M'$ and $-M''$ as a function of $I_{ac}$ at 1.7 K and $H = H_M$. At low $I_{ac}$ both $M'$ and $-M''$ vary smoothly. However, at $I_{ac} \sim 8.8$ mA, both $M'$ and $-M''$ exhibit a sharp jump followed by a series of further jumps at higher excitation currents. The VME associated with these jumps becomes apparent when we compare similar measurements performed at different $H$. Fig. 3(b) shows the variation of $-M''$ as a function $I_{ac}$ measured at different $H$ between 1.18 kOe to 4 kOe. We observe that the $I_{ac}$ at which the first jump occurs is strongly dependent on magnetic field: The jump occurs at the smallest value of $I_{ac}$ when $H \approx H_m$ and shifts to higher values as one moves away from $H_m$. Finally, with increase in temperature the sharp jumps are gradually smeared, and only a smooth variation is seen at above 2 K (Fig. 3(c)).

To understand the origin of these jumps, we note that for low $I_{ac}$, the induced $J_s^{ac}(r)$ will cause small alternating compression and rarefaction of the VL, similar to the elastic response of a solid under oscillating stress. The gradual change in $M'$ and $-M''$ in this regime reflects the anharmonicity of the pinning potential. As $I_{ac}$ is increased to a critical value, for some range of $r$, $J_s^{ac}(r)$ will reach the critical current density, $J_c$, and the vortices in this region will get delocalized from antidot. Above this value (which we denote as $I_{ac}^p$) within a circular annulus towards the periphery of the sample, the VL will start sliding over the antidot array. Thus the sharp increase in $M'$ signifies the threshold where the elastic continuity of the VL is destroyed by the sliding. Since each hop of a vortex from one antidot to the next will induce a $2\pi$ phase slip in the intervening superconducting barrier, such sliding motion will also cause the dissipation ($-M''$) to abruptly increase. The subsequent jumps in $M'$ and $-M''$ with increasing $I_{ac}$ can be understood as follows:

At $I_{ac}^p$ the vortex lattice over a circular annulus will slide back and forth over one lattice constant, $a$, of the antidot array in each cycle of a.c. excitation. However as $I_{ac}$ is increased two things will happen. First, the circular annulus over which the vortices are delocalized will gradually increase. Secondly, the amplitude of the oscillatory displacement of the vortices will increase. Thus subsequent jumps correspond to the excitation current values (denoted as $I_{ac}^1$, $I_{ac}^2$, …) where VL start to slide over higher integral multiples of the lattice constant ($2a$, $3a$,…). Also, it is interesting to note that the dissipative response shows shallow minima in between successive jumps. This shows that in addition to uniform motion of the VL, other mechanisms of strain relaxation not involving the entire delocalized VL region, such as sliding of vortices in one domain of the antidot over another creating lines of doubly occupied antidots or the fracture of the VL into disconnected icebergs of VL domains might also be playing a role. This is also responsible for the successive jumps to be gradually smaller in magnitude.

We now investigate whether the jump at $I_{ac}^p$ can be understood in terms of conventional depinning of vortices from the underlying antidot array. For this, we need to compare the depinning force ($F_p$) for an isolated vortex to overcome the pinning by an antidot with the force exerted on a vortex by the supercurrent ($F_{ac}^c$) at $J_s^{ac} \sim J_c$. We first calculate $J_s^{ac}(r)$ for the geometry of our coil and the sample. This can in principle be done by numerically solving the Maxwell equations and the London equation using the scheme outlined in ref. 22, by replacing the London penetration depth with the Campbell penetration depth[23,24] $\lambda_p$. However, here $\lambda_p$ is itself a function of $J_s^{ac}$, approximately following the semi-empirical relation[25], $\lambda_p(J_s^{ac}) = \frac{\lambda_p(0)}{(1-J_s^{ac}/J_c)^{1/4}}$. Therefore to calculate $J_s^{ac}(r)$ we adopt the following iterative procedure. Since $\lambda_p(J_s^{ac}) \approx \lambda_p(0)$ for $J_s^{ac} \ll J_c$, we use the value of $M'$ measured with $I_{ac}$ = 1 mA $\ll I_{ac}^p$ (for which $J_s^{ac}(r) \ll J_c$ over the entire sample) to obtain[26] $\lambda_p(0) \approx 3100\ nm$. Using this value of $\lambda_p(0)$ and a trial value of $J_c$, we incorporate the $J_s^{ac}(r)$ dependence of $\lambda_p$ for higher values of $I_{ac}$ and obtain a self-consistent current profile (see supplementary material[27]). $J_s^{ac}(r)$ increases linearly with $r$ for small $r$ and saturates beyond $r > 0.9$ mm, as shown in Fig. 4(a). As $I_{ac}$ is increased, this saturation current reaches $J_c$ at some critical $I_{ac}$, beyond which the iterative procedure fails to converge to a unique profile (Fig. 4(b)). This critical value $I_{ac} \approx I_{ac}^p$, where the VL on the outer annulus of the film ($r \gtrsim 0.9$ mm) starts sliding over the antidot array. The variational $J_c$ for our film is then chosen so that $I_{ac}^p$ is matched with the experimentally observed $I_{ac}^p$. This procedure yields $J_c \sim 1.2\text{-}1.4\times10^8$ A/m² at the first matching field for our sample, where the uncertainty is given by the difference between the maximum $J_s^{ac}(r)$ for which we obtain a stable solution and the trial $J_c$ value. Therefore, the force exerted on a single vortex for $J_s^{ac} \sim J_c$, is $F_{ac}^c = J_c \Phi_0 \sim 6.4 - 7.5 \times 10^{-15}$ N. To estimate $F_p$, we note that the confining potential created by the superconductor surrounding an antidot can be obtained from the vortex energy calculated in a thin film superconducting strip[28,29], i.e. $E(r) = \frac{\Phi_0^2}{2\pi\mu_0\Lambda} Ln\left(\frac{2w}{\pi\xi_{GL}}\sin\left(\frac{\pi(r-\frac{d}{2})}{w}\right)\right)$, where $\Lambda \,(= 2\lambda^2/t)$ is the Pearls penetration depth, $w$ is the width of the superconductor separating two antidots, and $\mu_0$ is the vacuum permeability. This relation is valid when the core of the vortex is completely inside the superconductor, i.e. $r > 2\xi_{GL}+d/2$. The form of the potential implies that the restoring force on the vortex will monotonically decrease as the vortex position is changed from the edge towards the center of the superconducting strip. Thus the depinning force is given by, $F_p = \frac{\partial E}{\partial r}\Big|_{r\sim(2\xi_{GL}+\frac{d}{2})} \sim \frac{\Phi_0^2}{4\pi\mu_0\Lambda\xi_{GL}} = 4.41 \times 10^{-14}$ N, where $\xi_{GL} \sim 8\,nm$ [17], which is one order of magnitude larger than $F_{ac}^c$. This difference cannot be accounted for by thermally activated hopping of vortices from the following energetic consideration. The potential barrier a vortex has to overcome to hop from one antidot to the adjacent one is given by, $E(r)|_{max} = \frac{\Phi_0^2}{2\pi\mu_o\Lambda} Ln\left(\frac{2w}{\pi\xi_{GL}}\right) \sim 7~meV$. On the other hand at 1.7 K, $k_BT \sim 0.15$ meV, such that thermally activated hopping will play a negligible role. Therefore, the delocalization of the vortices at $I_{ac}^p$ is clearly a collective phenomenon driven by an interplay of inter-vortex interaction and localization[30].

We now focus on the role of VME on $I_{ac}^p$. $I_{ac}^p$ is lowest for $H \approx H_m$ when the VL the most rigid. Physically this can be understood from the fact that the delocalization of vortices is governed both by the individual pinning potential of the antidot and the inter-vortex interaction. Away from the matching field the antidot array either contains some empty sites (for $H < H_M$) or sites that are doubly occupied by vortices (for $H > H_m$) which can accommodate change of flux through local rearrangement of vortices without breaking overall elastic continuity. Thus the VL gets collectively delocalized only when these local rearrangements become energetically unfavorable compared to the sliding of the entire VL. In contrast at $H_m$ where each antidot is filled with exactly one vortex, any such local rearrangement will trigger a collective delocalization of the entire VL. Subsequent thresholds, $I_{ac}^1$ and $I_{ac}^2$ also show similar variation with $H$, governed by similar energetic consideration.

It is now instructive to look at the evolution of $M'$ and $M''$ with $H$ for different $I_{ac}$. Fig. 5 (a)-(b) show $M'$-$H$ and (-$M''$)-$H$ at 1.7K corresponding to $I_{ac}$ values marked by vertical dashed arrows in Fig 3(a), while the magnetic field is swept from 10 to -10 kOe and back. For $I_{ac} < I_{ac}^p \sim 8.8~mA$, $M'$-$H$ traces remain qualitatively similar, and display minima at the matching fields. (-$M''$)-$H$ also displays shallow dips at the matching field. Both these are consistent with the elastic regime, where the VL becomes more rigid at $H_m$ and therefore each vortex oscillates with a smaller amplitude. Since no vortex hop is involved in this range -$M''$ is dominated by the Bardeen-Stephen dissipation due to the oscillating vortex in the superconducting region surrounding every antidot.

This behavior drastically changes at $I_{ac} = 9$ mA $\sim I_{ac}^p$. Here $M'$ develops small peaks at $H_m$ showing that the VL gets delocalized at the matching field. This leaves pronounced signature in the dissipative response and -$M''$ exhibits pronounced peaks at $H = H_m$. However, since this excitation cannot delocalize the VL away from the matching field, the elastic response is restored as we move away from $H_m$. As $I_{ac}$ is increased to 10 mA (midway between $I_{ac}^p$ and $I_{ac}^1$), the vortices are delocalized even away from $H_m$ and the peaks in $M'$ shifts to higher fields. At the same field -$M''$ shows an abrupt jump. Interestingly below the delocalization field, $M'$ shows minima at $H_m$ characteristic to the elastic response of the VL, whereas -$M''$ remains flat and featureless. To understand this difference, we note that when the outer annulus of the VL gets decoupled from the antidot array, the central region continues to behave as an elastic solid. This elastic core will contribute to exhibit matching effect in $M'$ whereas the outer region sliding over the antidot array will contribute a nearly field independent background. On the other hand since the central elastic core will contribute very little to dissipation, -$M''$ will be dominated by the delocalized portion of the VL in the outer annulus. The nearly constant -$M''$ shows that at this excitation the VL in the outer annular region remains delocalized over the entire range of fields between $H = \pm H_m$. We observe that this qualitative behavior repeats between $I_{ac}^1$ and $I_{ac}^2$, and, $I_{ac}^2$ and $I_{ac}^3$, though the matching effect in $M'$ becomes progressively weaker as larger fraction of the VL gets delocalized.

One interesting aspect of our data is the behavior at $H = 0$. The zero field minimum in $M'$ also transforms into a small peak with increase in $I_{ac}$, similar to the first matching field which is unexpected if we consider the antidot array as empty. In fact –$M'$-$I_{ac}$ also shows characteristic jumps[27] at $H = 0$ though the structure is more complex than at $H = H_m$. The most likely explanation for this observation is that the zero field state is not empty but is populated with pairs of vortex and anti-vortex in the antidot array which are expected to spontaneously form in a 2-D superconductor. These vortex-antivortex pairs can also get delocalized under a.c. excitation, but its dynamics will be much more complex owing to the possibility of vortex-antivortex annihilation by the excitation field, and is currently not captured within our model. The existence of vortex-antivortex pairs in zero field in an antidot array has also been conjectured in an earlier study investigating transport current-stimulated depinning[14] of the VL, where the zero field state exhibited a behavior similar to that at matching fields.

## IV. Summary and outlook

In summary, we have shown that the Mott-like state of the VL realized in a superconducting films with antidot array (where each vortex is localized in an antidot) transforms at a characteristic a.c. magnetic field excitation to a metal-like state where the vortices are delocalized. This transformation happens well below the conventional depinning threshold and shows pronounced VME confirming that it is a collective phenomenon governed both by strength of the individual pinning centers and the intervortex interactions. From this standpoint, our experimental observation is analogous to the dynamical vortex Mott insulator to metal transition induced by a transport current reported in ref. 13. However, in our case further investigation are required to understand whether this represents a true dynamical phase transition and why the characteristic jumps in *M* associated with this transition disappear well below $T_c$.

It is also interesting to note in this context, that our observation that $I_{ac}^p$ (and hence $J_c$) is lowest at the matching field, is different from the variation of critical current very close to $T_c$ [5], where the critical current was reported to show maxima at matching field. The origin of this difference can be traced to Little-Parks like quantum interference (QI) effects[31,32]. When the width of the superconductor (*w*) in the antidot array is smaller than $\xi_{GL}$, QI causes the super-current around each loop to go to zero at matching fields. Thus the amplitude of the superconducting order parameter which follows the variation in supercurrent gets enhanced at matching field, which in turn reflects as maxima[33] in $T_c$ and critical current. Therefore, QI will produce an effect opposite to VME on the critical current. Since $\xi_{GL}$ diverges close to $T_c$, QI will always dominate at these temperatures and give rise to maxima in critical current at matching fields. In contrast, our experiments are carried out at low temperatures where $\xi_{GL} \ll w$, and QI plays an insignificant role. Therefore, in addition to avoiding the effect of thermally excited vortex motion, our experiments are significant from the standpoint that they allows us to look at VME, uncontaminated by QI.

*Acknowledgements:* We thank Valerii Vinokur for critically reading the paper and for explaining the possible connection between our depinning transition and Mott-vortex to metal transition. We thank Bhagyashree A Chalke and Rudheer Bapat for help with scanning electron microscopy measurements. RS acknowledges useful discussions with Ahana Chakraborty. PR acknowledges Department of Atomic Energy, Government of India, and Science and Engineering Research Board, Government of India for financial support (Grant No. EMR/2015/000083). SB acknowledges partial financial support from the Department of Science and Technology, India through (No. SERB/F/1877/2012, SERB/F/745/2014) and the Indian National Science Academy through (SP/YSP/73/2012/1875).

[1] M. J. Higgins, and S. Bhattacharya, *Varieties of dynamics in a disordered flux-line lattice*. Physica C **257**, 232 (1996).

[2] Brandt, E. H. *The flux-line lattice in superconductors*. Rep. Prog. Phys. **58**, 1465–1594 (1995).

[3] V. V. Moshchalkov, M. Baert, V. V. Metlushko, E. Rosseel, M. J. Van Bael, K. Temst, Y. Bruynseraede, and R. Jonckheere, *Pinning by an antidot lattice: The problem of the optimum antidot size*. Phys. Rev. B **57**, 3615 (1998).

[4] U. Welp, Z. L. Xiao, J. S. Jiang, V. K. Vlasko-Vlasov, S. D. Bader, G. W. Crabtree, J. liang, H. Chik, and J. M. Xu, *Superconducting transition and vortex pinning in Nb films patterned with nanoscale hole arrays*. Phys. Rev. B **66**, 212507 (2002).

[5] A. V. Silhanek, L. Van Look, R. Jonckheere, B. Y. Zhu, S. Raedts, and V. V. Moshchalkov, *Enhanced vortex pinning by a composite antidot lattice in a superconducting Pb film.* Phys. Rev. B **72**, 014507 (2005).

[6] W. Vinckx, J. Vanacken, V. V. Moshchalkov, S. Matefi-Tempfli, M. Matefi-Tempfli, S. Michotte, and L. Piraux, *Vortex pinning in superconducting Nb thin films deposited on nanoporous alumina templates.* Eur. Phys. J. B **53**, 199 (2006).

[7] A. D. Thakur, S. Ooi, S. P. Chockalingam, J. Jesudasan, P. Raychaudhuri, and K. Hirata, Appl. *Vortex matching effect in engineered thin films of NbN.* Phys. Lett. **94**, 262501 (2009).

[8] G. R. Berdiyorov, M. V. Milošević, and F. M. Peeters, *Vortex configurations and critical parameters in superconducting thin films containing antidot arrays: Nonlinear Ginzburg-Landau theory.* Phys. Rev. B **74**, 174512 (2006).

[9] D. R. Nelson and V. M. Vinokur, *Boson localization and correlated pinning of superconducting vortex arrays,* Phys. Rev. B **48**, 13060 (1993).

[10] K. Harada, O. Kamimura, H. Kasai, T. Matsuda, A. Tonomura, V. V. Moshchalkov, *Direct Observation of Vortex Dynamics in Superconducting Films with Regular Arrays of Defects*, Science **274**, 1167 (1996).

[11] M. Baert, V. V. Metlushko, R. Jonckheere, V. V. Moshchalkov, and Y. Bruynseraede, *Composite Flux-Line Lattices Stabilized in Superconducting Films by a Regular Array of Artificial Defects.* Phys. Rev. Lett. **74**, 3269 (1995).

[12] S. Goldberg, Y. Segev, Y. Myasoedov, I. Gutman, N. Avraham, M. Rappaport, E. Zeldov, T. Tamegai, C. W. Hicks, and K. A. Moler, *Mott insulator phases and first-order melting in $Bi_2Sr_2CaCu_2O_{8+\delta}$ crystals with periodic surface holes.* Phys. Rev. B **79**, 064523 (2009).

[13] N. Poccia, T. I. Baturina, F. Coneri, C. G. Molenaar, X. R. Wang, G. Bianconi, A. Brinkman, H. Hilgenkamp, A. A. Golubov, V. M. Vinokur, *Critical behavior at a dynamic vortex insulator-to-metal transition.* Science **349,** 1202 (2015).

[14] Z. Jiang, D. A. Dikin, V. Chandrasekhar, V. V. Metlushko and V. V. Moshchalkov, *Pinning phenomena in a superconducting film with a square lattice of artificial pinning centers.* Appl. Phys. Lett. **84**, 5371 (2004).

[15] S. Kumar, C. Kumar, J. Jesudasan, V. Bagwe, P. Raychaudhuri, and S. Bose, *A two-coil mutual inductance technique to study matching effect in disordered NbN thin films.* Appl. Phys. Lett. **103**, 262601 (2013).

[16] S. Kumar, C. Kumar, J. Jesudasan, V. Bagwe, P. Parab, P. Raychaudhuri and S. Bose, *Origin of matching effect in anti-dot array of superconducting NbN thin films.* Supercond. Sci. Technol. **28**, 055007 (2015).

[17] M. Mondal, M. Chand, A. Kamlapure, J. Jesudasan, V. C. Bagwe, S. Kumar, G. Saraswat, V. Tripathi and P. Raychaudhuri, *Phase diagram and upper critical field of homogeneously disordered epitaxial 3-dimensional NbN films.* J. Supercond Nov Magn **24**, 341 (2011).

[18] M. Mondal, A. Kamlapure, M. Chand, G. Saraswat, S. Kumar, J. Jesudasan, L. Benfatto, V. Tripathi and P. Raychaudhuri, *Phase fluctuations in a strongly disordered s-wave NbN superconductor close to the metal-insulator transition.* Phys. Rev. Lett. **106,** 047001 (2011).

[19] M. Chand, G. Saraswat, A. Kamlapure, M. Mondal, S. Kumar, J. Jesudasan, V. Bagwe, L. Benfatto, V. Tripathi and P. Raychaudhuri, *Phase diagram of a strongly disordered s-wave superconductor, NbN, close to the metal-insulator transition.* Phys. Rev. B **85**, 014508 (2012).

[20] The Fourier transform is obtained using the image processing module of WSXM software: I. Horcas, R. Fernández, J. M. Gómez-Rodríguez, J. Colchero, J. Gómez-Herrero and A. M. Baro, *WSXM: A software for scanning probe microscopy and a tool for nanotechnology.* Rev. Sci. Instrum. **78**, 013705 (2007).

[21] For a more detailed discussion on the magnetic field variation of M', see Supplementary material, Section III.

[22] S. J. Turneaure, E. R. Ulm, and T. R. Lemberger, *Numerical modeling of a two‐coil apparatus for measuring the magnetic penetration depth in superconducting films and arrays.* J. Appl. Phys. **79**, 4221 (1996).

[23] A. M. Campbell, *The response of pinned flux vortices to low-frequency fields.* J. Phys. C: Solid State Phys. **2**, 1492 (1969).

[24] E. H. Brandt, *Penetration of magnetic ac fields into type-II superconductors.* Phys. Rev. Lett. **67**, 2219 (1991).

[25] R. Prozorov, R. W. Giannetta, N. Kameda, T. Tamegai, J. A. Schlueter, and P. Fournier, *Campbell penetration depth of a superconductor in the critical state.* Phys. Rev. B **67**, 184501 (2003).

[26] The procedure for obtaining λ from M' is outlined in ref. 22. Also see, M Mondal, *Phase fluctuations in a conventional s-wave superconductor: Role of dimensionality and disorder.* Ph. D. thesis, Tata Institute of Fundamental Research. (arXiv:1303.7396).

[27] Details of the computational procedure for obtaining $J_s^{ac}(r)$ is given in Section I of the supplementary material. Section II shows a comparison of $-M''$ as a function of $I_{ac}$ at $H = 0$ and $H = H_m$.

[28] V. G. Kogan, *Pearl's vortex near the film edge.* Phys. Rev. B **49**, 15874 (1994).

[29] J. R. Kirtley, C. C. Tsuei, V. G. Kogan, J. R. Clem, H. Raffy, and Z. Z. Li, *Fluxoid dynamics in superconducting thin film rings.* Phys. Rev. B **68**, 214505 (2003).

[30] The same conclusion can be arrived at by considering the upper bound of the vortex displacement due to a.c. magnetic field; see Supplementary Material, Section IV.

[31] W. A. Little and R. D. Parks, *Observation of Quantum Periodicity in the Transition Temperature of a Superconducting Cylinder.* Phys. Rev. Lett. **9**, 9 (1962).

[32] Y-L. Lin and F. Nori, *Quantum interference in superconducting wire networks and Josephson junction arrays: An analytical approach based on multiple-loop Aharonov-Bohm Feynman path integrals.* Phys. Rev. B **65**, 214504 (2002).

[33] M. J. Higgins, Yi Xiao, S. Bhattacharya, P. M. Chaikin, S. Sethuraman, R. Bojko, and D. Spencer, *Superconducting phase transitions in a kagomé wire network.* Phys. Rev. B **61**, R894(R) (2000).

**Figure Captions**

**Figure 1.** (a) Scanning electron micrograph showing the NbN film with antidot array. The panels on the right show the Fourier transforms of the areas bounded in the respective boxes. The six spots in the Fourier transform shows local hexagonal ordering. (b) Schematic diagram of the two-coil mutual inductance setup. (c) Schematic diagram showing the circulating a.c. supercurrent set up in a circular film with antidot array due to the alternating magnetic field from the primary coil. The width of the circles is proportional to the amplitude of $J_S^{ac}$. (d) Schematic description of the compression and rarefaction of the VL in one cycle of the alternating magnetic field. The leftmost panel shows the VL without external a.c. field.

**Figure 2.** Variation of $M'$ as a function $H$ at different temperatures with $I_{ac} \sim 5$ mA. VME manifest as minima in $M'$ at matching fields.

**Figure 3.** (a) $M'$ and $-M''$ as a function of $I_{ac}$ for fixed value of d.c. magnetic field $H = H_m = 2.37$ kOe at 1.7 K. Both quantities show a series of jumps above $I_{ac} \sim 8.8$ mA. The vertical dashed arrows correspond to $I_{ac}$ values for which $M'$-$H$ and -$M''$-$H$ are plotted in Figure 5. (b) -$M''$ as a function of $I_{ac}$ for different $H$ below and above $H_m$ at 1.7 K; for clarity each successive curve for $H > 1.18$ kOe is shifted upwards by 2 nH. The grey curve show the locus of $I_{ac}^p$. (c) -$M''$ as a function of $I_{ac}$ for $H = 2.37$ KOe at different temperatures; above 2 K the sharp jumps transform into a smooth variation.

**Figure 4.** (a) Simulated value of $J_S^{ac}$ along the radius of the film for $I_{ac} \sim 1 - 8.7$ mA. (b) Simulated value of $J_S^{ac}$ along the radius of the films for $I_{ac} \sim 8.8$ mA for the 14$^{th}$ and 15$^{th}$ iteration cycle; here the simulation does not converge to a unique value and the solution becomes bistable. The profile of the a.c. magnetic field ($B_{ac}$) arising from the primary coil is shown in the same panel.

**Figure 5.** (a) *M′-H* and (b) *-M′′-H* at different $I_{ac}$ at 1.7 K. The values of $I_{ac}$ correspond to the vertical dashed arrows shown in Fig. 3(a). The data is recorded by sweeping *H* from 10 kOe to -10 kOe and back. A small hysteresis between positive and negative field sweeps is observed in all curves. In panel (b) some of the curves have been shifted upward for clarity, by an amount mentioned next to the curve.

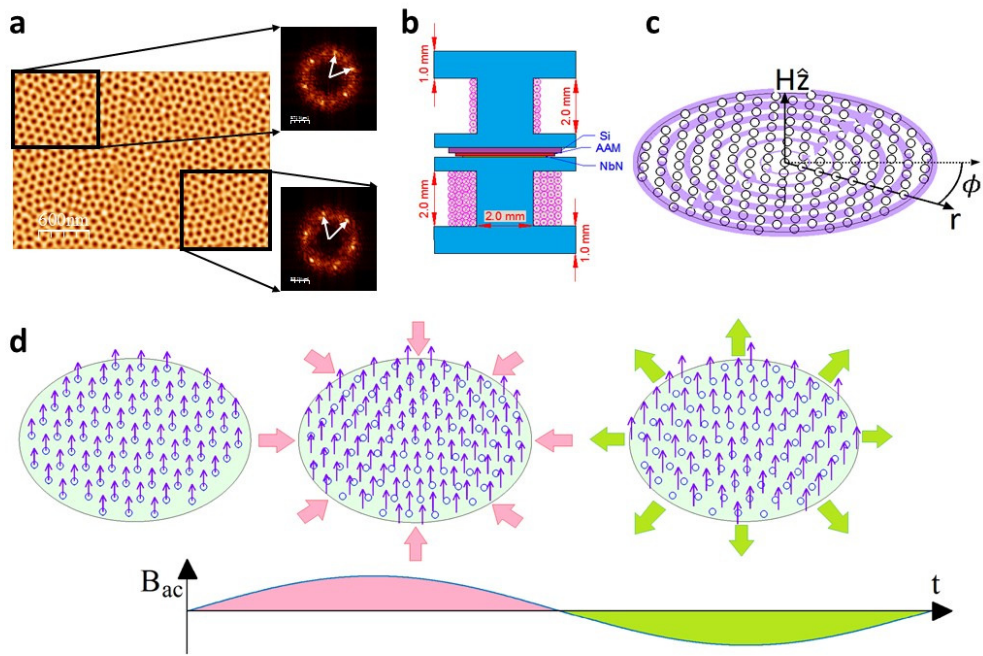

**Figure 1**

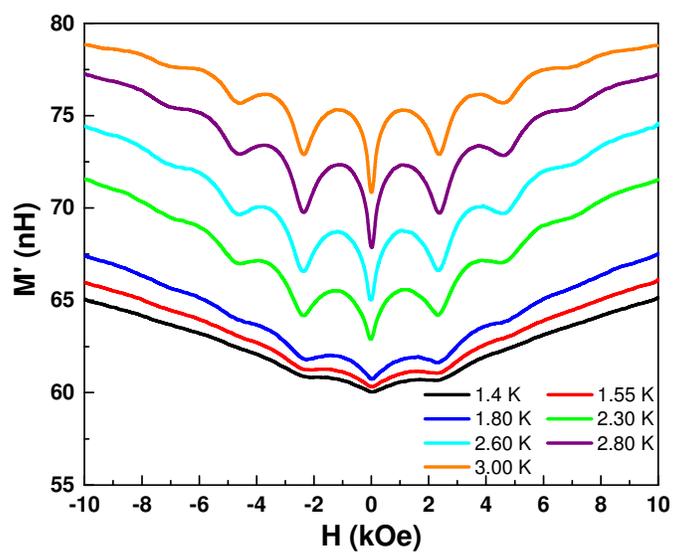

**Figure 2**

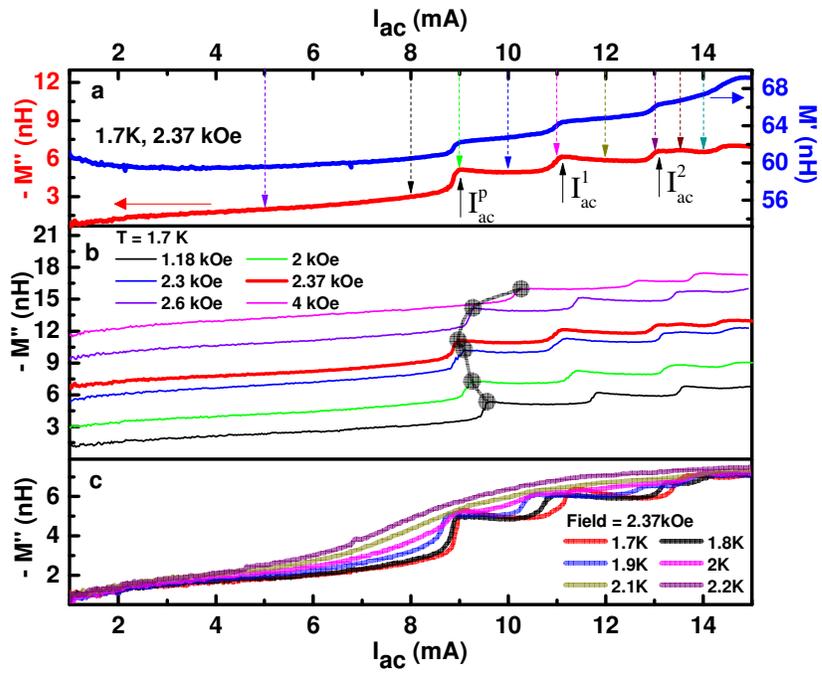

**Figure 3**

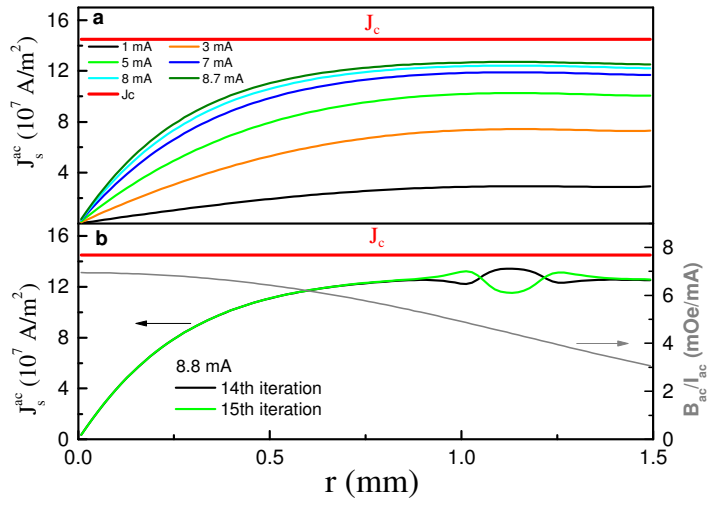

**Figure 4**

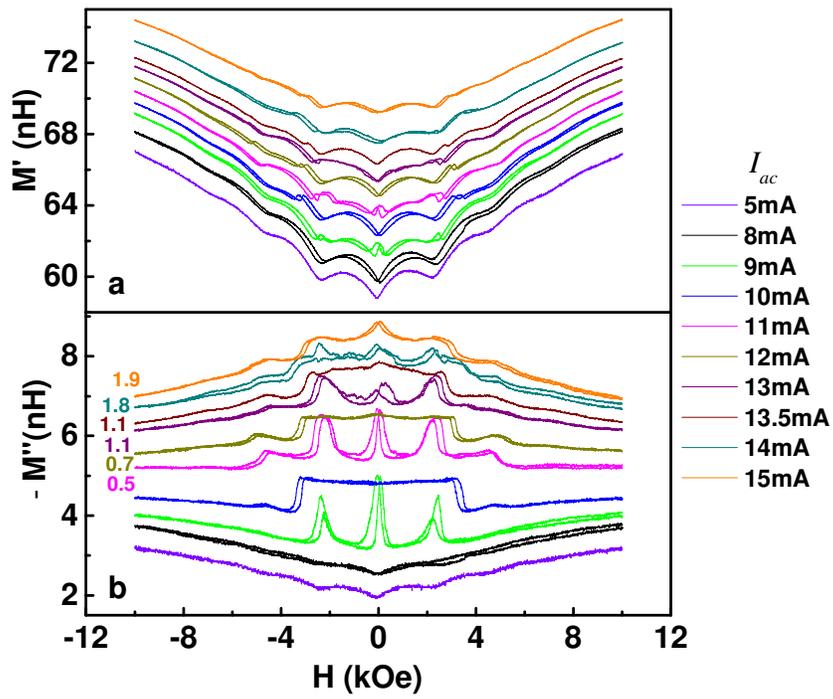

**Figure 5**

## Supplementary Material

### I. Calculation of $J_s^{ac}(r)$

By solving coupled Maxwell's equations and London equation for the geometry of our system we can in general obtain supercurrent density as a function of radial distance from the centre of the sample ($r$) for a constant value of penetration depth, $\lambda$. $J_s^{ac}(r)$ is obtained from the following self-consistent equations,

$$\boldsymbol{A}_{film}(\boldsymbol{r}) = \boldsymbol{A}_{drive}(\boldsymbol{r}) + \frac{\mu_0}{4\pi} \int d^3 r' \frac{J_s^{ac}(r')}{|r-r'|} \qquad (1)$$

$$J_s^{ac}(\boldsymbol{r}) = -\frac{A_{film}(r)}{\mu_0 \lambda^2}, \qquad (2)$$

where the vector potential on the films, $A_{film}(r)$, is a combination of the vector potential resulting from the current passing in the drive coil, $A_{drive}(r)$, and the induced supercurrent density on the film. The same equations can be inverted to obtain the value of $\lambda$ from the experimental value of $M'$. The scheme for numerically solving these equations for a film with circular geometry is identical to that used in refs. 1 and 2 is described in detail in refs. 3 and 4.

In our system the additional complicacy arises from the fact that Campbell penetration depth, $\lambda_p$, is a function of the local supercurrent density, $J_s^{ac}(r)$ and critical current density, $J_c$ following a quasi-empirical relation,

$$\lambda_p(J_s) = \frac{\lambda_p(0)}{(1-J_s^{ac}/J_c)^{1/4}} \qquad (3)$$

Therefore, to obtain $J_s^{ac}(r)$ we had to adopt an iterative procedure. First, using the value of $M'$ for $I_{ac} = 1\text{mA} \ll I_{ac}^p$, we obtain the value of $\lambda_p(0) \approx 3100\ nm$ using standard procedure[1,2]. For this $I_{ac}$ value $J_s^{ac}(r) \ll J_c$ for all values of $r$ and therefore $\lambda_p$ is assumed to be constant over the entire film, i.e. $\lambda_p(J_s^{ac}) \approx \lambda_p(0)$.

To obtain the $J_s^{ac}(r)$ at higher currents, we first note that the depinning at matching field occurs at $I_{ac} \sim 8.8\ mA$; at this value $J_s^{ac}(r) \sim J_c$ on the outer circular annulus of the sample. At this excitation it should theoretically not be possible to obtain $J_s^{ac}(r)$ over the entire film since eqn. 3 would no longer remain valid. Therefore we take a value just below, i.e. $I_{ac} \sim 8.7\ mA$ and first obtain $J_s^{ac}(r)$ as a function of $r$ using constant value of $\lambda_p = \lambda_p(0)$. Now to invoke the $J_s^{ac}(r)$ dependence of $\lambda_p$, we choose a trial value for $J_c$ as $2.65 \times 10^8 A/m^2$, which is marginally larger than the maximum value of $J_s^{ac}(r)$ (typically $J_s^{ac}(r = 1.5\ mm)$) obtained from the previous constant-$\lambda_p$ simulation. This is done so as to keep the value of $\lambda_p$ obtained from eqn. 3 real in all steps during the iteration. Using this value of $J_c$ and $J_s^{ac}(r)$ we calculate $\lambda_p(r)$. Then using the obtained $\lambda_p(r)$ refine the value of $J_s^{ac}(r)$ and keep repeating this process iteratively. We observe that this iterative procedure leads to an oscillatory convergence of $J_s^{ac}(r)$ (Fig. S1(a)). To achieve a faster convergence we therefore take the average of the values of $\lambda_p(r)$ in the last two steps of the iteration and solve to obtain the new values for $J_s^{ac}(r)$. This modified algorithm leads to a faster convergence as shown in Fig. S1(b).

We observe that, as expected, at small $r$ the final value of $J_s^{ac}(r)$ is similar to the value calculated for constant $\lambda_p$ but becomes significantly smaller at larger values of $r$. Therefore to obtain $J_c$ for our sample we again take a new trial value which is slightly higher than the maximum value of $J_s^{ac}(r)$ obtained from the last iteration. Then we run the iterative procedure again, starting from with $\lambda_p(r)$ obtained from the last step of the previous iteration, to obtain $J_s^{ac}(r)$ for this new trial $J_c$. When this step is repeated several times we observe that below $J_c \sim 1.45 \times 10^8 A/m^2$ the iterative procedure fails to converge. This signals the depinning threshold where eqn. 3 is no longer applicable. We observe that our iterative procedure converges till the maximum value of $J_s^{ac}(r)$ is within 20% of $J_c$.

To cross-check the consistency of the procedure we fix $J_c \sim 1.45 \times 10^8\ A/m^2$ for which we get a stable solution for $I_{ac} \sim 8.7$ mA and run the simulation from the other end by gradually increasing the value of $I_{ac}$ from 1mA in steps of 0.1 mA and run the iteration for each value. For each $I_{ac}$ we start the iteration using the converged value of $\lambda_p(r)$ obtained for the previous value of $I_{ac}$. This gives us $J_s^{ac}(r)$ for different $I_{ac}$ (Fig. S1(c)). At $I_{ac}$ = 8.8 mA the iterative procedure fails to converge (Fig. S1(d)). We thus conclude that this is the physical value of $J_c$ in our film with an error of about 20%.

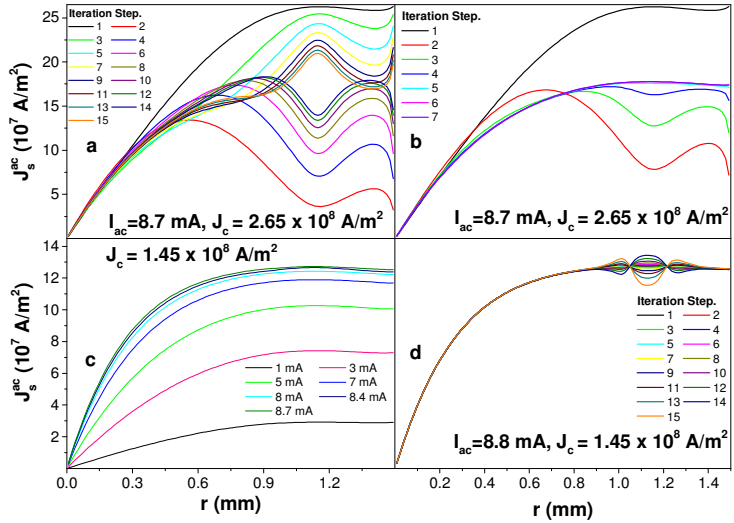

**Fig S1.** (a) Calculated value of $J_s^{ac}(r)$ at each step of iteration starting with a constant $\lambda_p$ (Step 1). We observe a slow oscillatory convergence. (b) Calculated values of $J_s^{ac}(r)$ at each step of iteration using the modified iterative procedure. Here the iteration converges after the 6$^{th}$ step. (c) Calculated values of $J_s^{ac}(r)$ for different $I_{ac}$ using the modified iterative procedure. (d) Calculated values of $J_s^{ac}(r)$ at each step of iteration for $I_{ac}$ = 8.8 mA using the modified iterative procedure. Here the iteration fails to converge for $r > 0.9$ mm indicating that $J_s^{ac}(r) \sim J_c$.

## II. -M″ as a function of $I_{ac}$ at zero field and first matching field

In Figure 5 of the main paper we have shown that $M'$ and $M''$ at zero field follows an evolution that is similar to that at matching fields. Here we compare -$M''$ as a function of $I_{ac}$ at zero field and at the first matching field on a different sample but grown on a similar anodized alumina template. We observe jumps both at $H = 0$ and $H = H_m$. However, the data at $H=0$ show an additional smaller jump (marked by arrow), not seen at $H = H_m$.

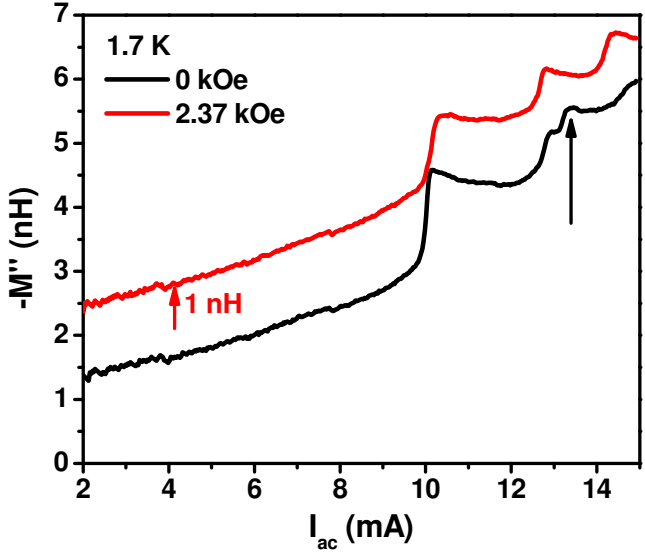

**Fig S2.** -M″ as a function of $I_{ac}$ for fixed value of d.c. magnetic field H = 0 kOe and H= $H_m$ = 2.37 kOe at 1.7 K. The curve for 2.37 kOe is shifted upwards by 1 nH for clarity. The data for 0 kOe shows an extra jump shown by the black arrow.

## III. Screening response from Meissner supercurrent and Oscillation of Vortices

In our experiment, the screening response in the presence of vortices arises from two sources: The screening of the magnetic field from Meissner supercurrents and the periodic compaction and rarefaction of the vortex lattice due to the a.c. magnetic field. The effective penetration depth is thus given by, $\lambda^2(H) = \lambda_M^2 + \lambda_v^2(H)$, where $\lambda_M$ is from Meissner supercurrent and $\lambda_v$ is from the motion of vortices. These two penetration depths are analogous to the London penetration depth and the Campbell penetration depth[5] in a bulk Type II superconductor. Fig **S3** shows the magnetic field variation of $\lambda(H)$ at 1.7 K measured with $I_{ac}$ = 5 mA. If we assume that *to the lowest order*, the vortex lattice is empty in zero field then, $\lambda_M \approx \lambda(0) = 3009\ nm$. On the other hand at the first matching field $\lambda(H = H_M) \approx 3144\ nm$. This gives $\lambda_v(H = H_M) \approx 911 nm$, which is the vortex contribution to the penetration depth at this field. Therefore the dominant response to the shielding response is always from the Meissner current, which accounts for the relatively small amplitude variation in $M'$ over the large background.

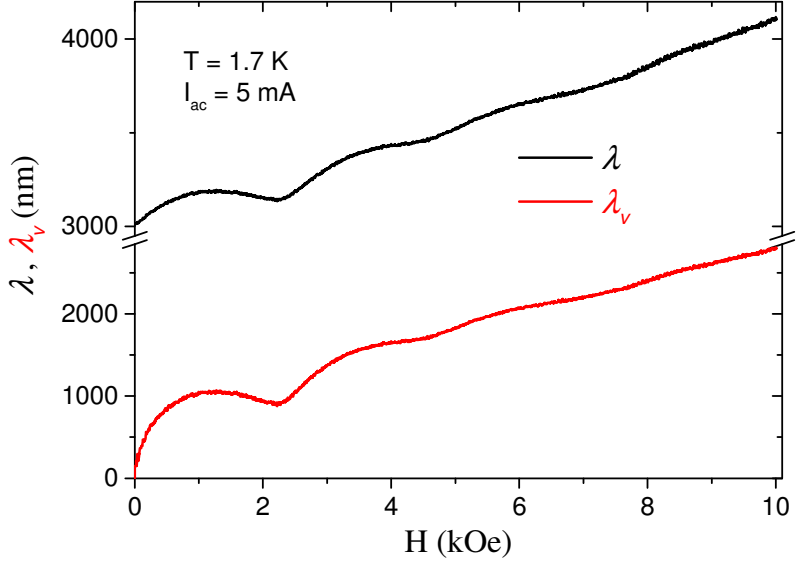

**Fig S3.** Magnetic field variation of $\lambda$ obtained from $M'$ and $\lambda_v = \sqrt{\lambda(H)^2 - \lambda_M^2}$ at 1.7 K.

In this context, it is also instructive to analyze the magnetic field variation of $\lambda_v$ between $H = 0$ and $H = H_m$. $\lambda_v$ which arises from this oscillatory motion of the vortex inside a potential well is given by[5], $\lambda_v^2(B) = \frac{\Phi_o B}{4\pi\alpha}$, where $\alpha$ is the Labusch parameter[6] which is proportional to the restoring force on the vortex when it is moved from its equilibrium position. At very small magnetic field, vortices are far from each other and $\alpha$ is determined essentially by the field independent restoring force arising from the antidot. In this regime $\lambda_v^2(B)$ increases with increasing $B$. However, as the magnetic field is increased, very soon the vortices start experiencing the repulsive interaction from all surrounding vortices, which further confines the vortex in its equilibrium position, thereby increasing $\alpha$. Thus $\alpha$ increases monotonically with field, and attains its maximum value at the first matching field, where each vortex is confined in a cage formed by the vortices in neighboring antidots. Thus, the variation of $\lambda_v^2(B)$ is governed by a combination of two effects: (i) an increase with $B$ due to the $B$ appearing in the numerator of the expression and (ii) a decrease with $B$ due to the increase $\alpha$ appearing in the denominator. As a result $\lambda_v^2(B)$ first increases with field goes through a maxima and somewhere midway between 0 and $H_m$ and then exhibits a minima at $H_m$ where α is maximum. Consequently, $M'$, mimics the same variation.

**IV. Upper bound of vortex displacement from its mean position at the matching field due to the a.c. magnetic field**

In this section we estimate an upper bound of the displacement of a vortex from its mean position due to the a.c. magnetic field. Assuming that pinning is infinitesimally small such that the entire a.c. field get through, the vortex lattice constant would be given by $a_v = \left(\frac{2\Phi_0}{\sqrt{3}B}\right)^{1/2}$, where $B=B_{dc}+B_{ac}\sin(\omega t)$. The role of the a.c. field is to introduce periodic variation of $a_v$ centered about its value for $B = B_{dc}$. Since the vortices enters and exit from the periphery of the sample (see Fig. 1d), the vortex at the center of the sample remains at rest whereas the remaining of the vortices undergo periodic oscillations along the radial direction whose amplitude linearly increases as we go towards the periphery. Thus for a sample with radius R, the maximum displacement from its mean position experienced by a vortex located close to the periphery of the sample is given by, $D_{max}=(R/a_v)*\Delta a_v$, where $\Delta a_v = \left(\frac{2\Phi_0}{\sqrt{3}B_{dc}}\right)^{1/2} - \left(\frac{2\Phi_0}{\sqrt{3}(B_{dc}+B_{ac})}\right)^{\frac{1}{2}} \approx a_v\left(\frac{B_{ac}}{2B_{dc}}\right)$. At the first matching field, $a_v = a = 107$ nm, and for our sample, R = 1.5 mm. Therefore for $B_{ac} \sim 62$ mOe at which we observe the delocalization transition of the VL, $(D_{max}/a) \sim 0.2$. It is interesting to note that $D_{max}$ is less than half the lattice constant of the antidot, so that the a.c. field would not cause a vortex hop even for infinitesimally small pinning. In practice the pinning from the antidot would try to hold the vortex in place so the displacement would be much smaller. This is another way to understand why the Mott to metal transition is necessarily a collective phenomenon.

---

[1] A. Kamlapure, M. Mondal, M. Chand, A. Mishra, J. Jesudasan, V. Bagwe, L. Benfatto, V. Tripathi and P. Raychaudhuri, *Penetration depth and tunneling studies in very thin epitaxial NbN films.* Appl. Phys. Lett. **96**, 072509 (2010).

[2] M. Mondal, A. Kamlapure, M. Chand, G. Saraswat, S. Kumar, J. Jesudasan, L. Benfatto, V. Tripathi and P. Raychaudhuri, *Phase fluctuations in a strongly disordered s-wave NbN superconductor close to the metal-insulator transition.* Phys. Rev. Lett. **106,** 047001 (2011).

[3] S. J. Turneaure, E. R. Ulm, and T. R. Lemberger, *Numerical modeling of a two-coil apparatus for measuring the magnetic penetration depth in superconducting films and arrays.* J. Appl. Phys. **79**, 4221 (1996).

[4] M Mondal, *Phase fluctuations in a conventional s-wave superconductor: Role of dimensionality and disorder.* Ph. D. thesis, Tata Institute of Fundamental Research. (arXiv:1303.7396).

[5] A. M. Campbell, *The response of pinned flux vortices to low-frequency fields.* J. Phys. C: Solid State Phys. **2**, 1492 (1969); *The interaction distance between flux lines and pinning centres.* **4**, 3186 (1971).

[6] R. Labusch, *Elasticity Effects in Type-II Superconductors.* Phys. Rev. **170**, 470 (1968).